\documentstyle[11pt,newpasp,twoside]{article}
\def\edcomment#1{\iffalse\marginpar{\raggedright\sl#1\/}\else\relax\fi}
\reversemarginpar

\begin{document}
\title{ULTRACAM -- an ultra-fast, triple-beam CCD camera} 

\author{Vik Dhillon$^1$, Tom Marsh$^2$ and the ULTRACAM team$^3$}
\setcounter{footnote}{3}\footnotetext{John Kelly, Richard Pashley, Mark 
Stevenson\\ {\em Department of Physics and Astronomy, University of 
Sheffield, Sheffield S3 7RH, UK}\\David Atkinson, Steven Beard, 
Derek Ives, Tully Peacocke, Chris Tierney, Andy Vick\\{\em UKATC, Royal 
Observatory, Blackford Hill, Edinburgh EH9 3HJ, UK}.}

\affil{$^1$Department of Physics and Astronomy, University of Sheffield,
Sheffield S3 7RH, UK\\
$^2$Department of Physics and Astronomy, Southampton University, 
Highfield, Southampton S017 1BJ, UK}

\begin{abstract}
ULTRACAM is an ultra-fast, triple-beam CCD camera which has been designed
to study one of the few remaining unexplored regions of observational 
parameter space -- high temporal resolution. The camera will see first
light in Spring 2002, at a total cost of $\pounds$300 k, and will be used
on 2-m, 4-m and 8-m class telescopes to study astrophysics on the fastest
timescales.
\end{abstract}

\section{Science}

The fastest timescale variations likely to be observed in an astrophysical 
environment are milliseconds, corresponding to the innermost orbits around 
neutron stars and black holes. Variations of faint sources on timescales 
longer than a few seconds have already been explored by conventional CCD 
instruments. ULTRACAM will explore the region of observational parameter 
space which lies between these two extremes, namely photometry of faint 
objects on timescales of seconds to milliseconds. The resulting scientific 
applications will include the study of pulsars, occultations, XRBs (e.g. 
searching for the optical analogue of the kilohertz QPOs), speckle imaging, 
precise timing studies (e.g. measuring eclipses), echo mapping, eclipse 
mapping of CVs, asteroseismology, and flickering and oscillations in CV 
accretion discs. 

The five essential requirements for such work are: 1. The capability of 
taking short exposures (from milliseconds to seconds) with essentially
no dead-time between exposures; 2. Multi-channels (3 or more) covering a wide
wavelength range ($u'$ to $z'$) in order to distinguish a blackbody spectrum 
from a stellar spectrum; 3. Simultaneous measurement of the different 
wavelength bands to avoid problems due to stochastic variations; 4. 
Imaging capability in order to simultaneously measure target, sky and 
comparison stars and to avoid the need for fixed apertures; 5. High 
efficiency and the capability of being mounted on various large aperture 
telescopes. These five requirements result in the instrument design presented 
here.

\section{Design}

\noindent{\bf Optics.} Light from the telescope is first collimated and then 
split into 3 bandpasses ($u'$, $g'$ and one of $r', i' or z'$) using 
dichroic beamsplitters and SDSS filters. The 
light from each filter is then re-imaged onto one of 3 
CCDs at a plate scale of 0.3 
arcseconds/pixel, giving a field of view of 5 arcminutes. This design ensures
that: 1. There is an 80\% probability of finding an R=12 magnitude comparison 
star in the field; 2. There is no ghosting due to the dichroics; 3. Only the 
collimator needs to be changed to mount ULTRACAM on a different telescope. 

\vspace*{0.1cm}

\noindent{\bf Mechanics.} The opto-mechanical chassis is a double-octopod 
constructed of aluminium plates and carbon fibre struts, giving a stiff, 
compact (75cm long), lightweight (75 kg) structure which is relatively 
insensitive to temperature variations. These characteristics make ULTRACAM 
highly portable and mountable on both small and large aperture telescopes. 

\vspace*{0.1cm}

\noindent{\bf Detectors.} ULTRACAM will use 3 Marconi (formerly EEV) 47-20 
AIMO, back-thinned, AR-coated CCDs. These will be Grade 0 devices (i.e. of 
the highest cosmetic quality) and contain 1024x1024 image pixels, each of
13 microns, with 1024x1024 masked pixels (i.e. they are frame-transfer 
devices). The chips exhibit exceptionally high QE (up to 97\%) and low
noise (3 e$^-$/pixel readout noise and $<0.1$ e$^-$/pixel/s dark current,
achieved by cooling to 233 K using a 3-stage Peltier stack and water 
cooling).

\vspace*{0.1cm}

\noindent{\bf Data acquisition and reduction.} Data from the 3 CCDs is first 
read by an SDSU controller and then passed to a dual-processor PC running 
RTLinux. Each exposure is time-stamped by a GPS receiver to an accuracy of 
0.01 milliseconds and then written to a RAID array and archived to tape. 
Pipeline data reduction software then fully reduces each data frame. Such a 
data acquisition and reduction system will enable us to observe for a whole 
night at the highest data rates (3.6 Mbytes/s) without stopping, building up 
light curves in real time. 

\section{Performance}

Frame transfer chips allow a completed exposure to be rapidly shifted 
from the imaging area to the masked area and then read out whilst the
next exposure is in progress. This, in conjunction with the high-speed 
data acquisition system described above, results in a full-frame readout
time of 5 milliseconds (as long as the exposure time is $>2$ s). By using
small windows and stacking the windows in the masked area (a procedure known
as `drift mode'), it is possible to obtain dead-times of 0.05 milliseconds
(as long as the exposure time is $>2$ milliseconds), thereby meeting
our requirement of millisecond exposure times with negligible dead time.

The optics and coatings in ULTRACAM have been designed with throughput
as the highest priority. This, in conjunction with the exceptional QE
performance of the CCDs, means that we expect to detect objects as
faint as $\sim 17$th magnitude at a signal-to-noise ratio of 3 in a 1
millisecond exposure on an 8-m class telescope.

\vspace*{0.25cm}

\noindent For more information on ULTRACAM, please consult the instrument's web-site
at http://www.shef.ac.uk/$\sim$phys/people/vdhillon/ultracam.

\end{document}